\documentclass[a4paper,12pt]{article}
\usepackage[round,authoryear]{natbib}
\usepackage{setspace,graphicx,multirow,sectsty,amsmath,amsthm,amssymb,latexsym,subeqnarray}
\usepackage[margin=10pt,font=small,labelfont=bf,labelsep=period,format=hang,textfont={footnotesize,sl}]{caption}

\hyphenchar\font=`\-
\renewcommand{\captionfont}{\sffamily}
\renewcommand{\captionlabelfont}{\small\sffamily}
\bibpunct[, ]{(}{)}{;}{a}{}{,}

\begin{document}
\bibliographystyle{oleb}
\begin{center}
{\large \bf Astrobiological Complexity with Probabilistic Cellular
Automata}

\vspace{0.5cm}

\large Branislav Vukoti\'c

\vspace{0.1cm}

\large Milan M.\ \'Cirkovi\'c

\vspace{0.1cm}

{\it Astronomical Observatory Belgrade, \\
Volgina 7, 11160 Belgrade-74, Serbia \\
E-mail: {\tt bvukotic@aob.rs}}

\vspace{0.7cm}
\end{center}

\begin{abstract}
\noindent The search for extraterrestrial life and intelligence
constitutes one of the major endeavors in science, but has yet
been quantitatively modeled only rarely and in a cursory and
superficial fashion. We argue that probabilistic cellular
automata (PCA) represent the best quantitative framework for
modeling the astrobiological history of the Milky Way and its
Galactic Habitable Zone. The relevant astrobiological parameters
are to be modeled as the elements of the input probability matrix
for the PCA kernel. With the underlying simplicity of the
cellular automata constructs, this approach enables a quick
analysis of large and ambiguous space of the input parameters. We
perform a simple clustering analysis of typical astrobiological
histories with "Copernican" choice of input parameters and
discuss the relevant boundary conditions of practical importance
for planning and guiding empirical astrobiological and SETI
projects. In addition to showing how the present framework is
adaptable to more complex situations and updated observational
databases from current and near-future space missions, we
demonstrate how numerical results could offer a cautious
rationale for continuation of practical SETI searches.
\end{abstract}


PACS number(s): 87.18.-h, 89.75.Fb, 89.90.+n, 02.70.-c

\section{Introduction: Why PCA in astrobiology?}

The early works of Stanislaw Ulam and John von Neumann in the
1940s, the book entitled ``Calculating space" by \citet{Zuse69},
John Conway's popular \textit{Game of Life} \citep[described
in][]{Gardner70}, and the first systematic analysis of Stephen
Wolfram in 1983-84 \citep{Wolfram83,Wolfram84}, with progressing
computational power over the past decades established a new
approach---or philosophy---in making scientific models of various
phenomena. Cellular automata (CA) modeling techniques are
increasingly gaining momentum in studies of complex systems and
their unpredictable behaviour. The CA operates on a lattice of
cells in discrete time steps. Each cell is characterized by a
state which evolves in time according to transition rules.
Transition rules define the state of the cell in the next time
step in relation to the present state of the cell itself and the
states of the cells in its surrounding
(neighbourhood)\footnote{In the theory of cellular automation,
the neighbourhood of a cell consists of the cell itself and the
surrounding cells. In this work we will use the Moore
neighbourhood and the term "surrounding" to refer to the
neighbourhood of the cell without the cell in question --
comprised of only the surrounding cells.}. Even simple transition
rules can result in a substantial complexity of emerging
behaviour \citep[for details on CA theory see][]{Ilachinski01}.
Despite the fact that deterministic CA can create some
random-like patterns, probabilistic cellular automata (PCA) are
more convenient tool to be used in discrete modeling of
intrinsically stochastic phenomena.

Probabilistic cellular automata have been studied extensively
\citep{BennettGrinstein85,GrinsteinJH85} and have shown good
results in practice as a lucrative modeling tool in many fields
of science and technology (e.g., modeling of forest fires,
pandemics, immune response, urban traffic, etc.
\citep{BattyCouclelisEichen97,HoyaWhite06,SoaresFilhoCerqueiraPennachin02,Torrens00}.
In particular, the application to biological sciences gives us a
better explanation of the microscopic mechanisms that lead to the
macroscopic behavior of the relevant systems
\citep{Borkowski09,deOliveira02,WoodAcklandLenton06}. These models
are simple and yet exhibit very intricate behavior -- not yet well
understood -- partly as a consequence of taking into account the
fluctuations that play an important role in determining the
critical behavior of the system considered. One important feature
of almost all these models is the presence of phase transitions,
which have potential to explain a wide variety of
phenomenological features of biological and ecological systems
\citep{Bak97,1997LNP...480..341B,Langton90,WoodAcklandLenton06}.
This is crucial for our attempt to extend the domain of numerical
simulations to astrobiology.

Astrobiology is a nascent multidisciplinary field, which deals
with the three canonical questions: How does life begin and
develop in its widest cosmical context? Does life exist elsewhere
in the universe? What is the future of life on Earth and in space?
\citep{desMarais99,Grinspoon03,ChybaHand05} A host of important
discoveries has been made during the last decade or so, the most
important certainly being the discovery of a large number of
extrasolar planets; the existence of many extremophile organisms,
some of which possibly comprise the "deep hot biosphere" of Thomas
Gold; others are living at altitude up to 41 km in the
stratosphere; the discovery of subsurface water on Mars and the
huge ocean on Europa, and possibly also Ganymede and Callisto;
the unequivocal discovery of many amino-acids and other complex
organic compounds in meteorites; modelling organic chemistry in
Titan's atmosphere; the quantitative treatment of the Galactic
habitable zone; the development of a new generation of panspermia
theories, spurred by experimental verification that even
terrestrial microorganisms easily survive conditions of an
asteroidal or a cometary impact; progress in methodology of the
Search for ExtraTerrestrial Intelligence (SETI) studies, etc. In
spite of all this lively research activity, there have been so far
surprisingly few attempts at building detailed numerical models
and quantitative theoretical frameworks which would permit an
understanding of the accumutating empirical data or for adding
rigor to the many hand-waving hypotheses which are thrown around.
Some of the excellent exceptions to this are studies of
Lineweaver and collaborators
\citep{Lineweaver01,LineweaverDavis02,Lineweaver04}, on the age
distribution of Earthlike planets and the structure of the
Galactic Habitable Zone \citep[GHZ; see][]{Gonzalez01}. It is
sometimes stated that we understand the underlying
"astrobiological dynamics" still so poorly; while that is
undoubtedly true, there have been well-documented cases in the
history of physical science (including the paradigmatical case of
neutron diffusion through a metallic shield which was
investigated by Ulam and von Neumann) which demonstrate that
various possible local dynamical behaviours converged toward
similar globally interesting physical outcomes. It is exactly this
motivation which prompts us to suggest the usage of PCA in
studying astrobiological complexity and to show that this can
offer us interesting, though unavoidably very preliminary,
insights.

(Some quantitative models have been developed in order to justify
or criticize particular SETI approaches. In general, they follow
one of the two schools of thought from the early 1980s, being
either (1) based on some extension of biogeography equations,
starting with \citet{Newman81}, and recently used by
\citet{Bjork07}; or (2) making use of discrete modeling, starting
with the work of \citep{Jones81}, and developed in rather limited
form by Landis, Kinouchi, and others
\citep{Landis98,Kinouchi01,CottaMorales09,Bezsudnov10}. The latter
studies used particular aspects of the discrete approach to
astrobiology, but have not provided a comprehensive grounding for
using such models rather than other, often more developed
numerical tools. In addition, they have of necessity been limited
by either arbitrary or vague boundary conditions and the lack of
specific astrophysical input dealing with the distribution of
matter in the Milky Way and possible risk factors. While we use
the results of the latter, "discrete" school of thought as a
benchmark \citep[in particular those of Cotta and
Morales][]{CottaMorales09}, we attempt to show how they could be
generalized to a wider scheme, encompassing not only SETI, but
much more general issues of astrobiological complexity.)

The plan of the paper is as follows. In the remainder of the
Introduction we review some of the motivations for a digital
perspective on astrobiology in general, and the usage of PCA in
particular. In Sec.~II our probabilistic model of astrobiological
complexity of the Galaxy is introduced and its main results
reviewed. In Sec.~III we discuss the key issue of boundary
conditions, especially in their relationship to Fermi's paradox
and biological contingency argument. Finally, in the concluding
section, we summarize our main results and indicate directions
for future improvement.

\subsection{Discrete nature of the distribution of matter}
While the present approach uses global symmetries of the Galactic
system (planarity, thin disk, thick disk, spiral arms, etc.), one
should keep in mind that the realistic distribution of baryonic
matter is discrete. In particular, stars possessing habitable
planets are hypothesized to form a well-defined structure, the
GHZ: an annular ring-shaped subset of the thin disk. Since the
lifeforms we are searching for, depend on the existence and
properties of habitable Earth-like planets, the separation of the
order of $\sim 1$ pc between the neighboring planetary systems
(characterizing the GHZ) ensures that, even if some exchange of
biologically relevant matter between the planetary systems occurs
before the possible emergence of technological, star-faring
species (as in classical panspermia theories), it remains a very
small effect, so the assumption of discrete distribution holds.
Even if an advanced technological civilization arises eventually
and engages in interstellar travel or decides to live in habitats
independent of Earth-like planets, it is to be expected that their
distribution will stay discrete for quite a long time, since the
resources necessary for interstellar travel (likely to be
expensive at all epochs) will remain distributed around Main
Sequence stars.

\subsection{Contingency in biological sciences}
The issues of determinism vs. indeterminism and contingency vs.\
convergence in biological sciences has been a very hotly debated
one ever since Darwin and Wallace published their theories of
evolution through natural selection in 1859. One of the main
opposing views has in recent years been put forward by proponents
of contingent macroevolution, such as \citet{Gould89,Gould96} or
\citet{McShea98}. According to this view, the contingent nature of
biological evolution guarantees that the outcome is essentially
random and unrepeatable. When this essential randomness is
coupled with the stochastic nature of external physical changes,
especially dramatic episodes of mass extinctions, we end up with
a picture where the relative frequency of whatever biological
trait (including intelligence, tool-making and other
pre-requisites for advanced technological civilization) is, in a
sufficiently large ensemble, proportional only to the relative
size of the relevant region of morphological space. While
proponents of this view do not explicitly mention astrobiology,
it is clear that the required ensemble can be provided only in
the astrobiological context \citep{Fry00}, notably by GHZ. (Of
course, the definition of morphological space hinges on the
common biochemical basis of life, although it does not seem
impossible to envision a generalization.)

On the  diametrically opposite end of spectrum, Conway Morris
\citep{ConwayMorris98,ConwayMorris03} argues that convergent
processes led to the current general landscape of the terrestrial
biosphere, including the emergence of intelligence in primates.
\citep{Dawkins89} and \citep{Dennett95} are certainly closer to
this position, though they are somewhat reserved with respect to
Conway Morris' unabashed anthropocentrism \citep[see
also][]{Sterelny05}. For some of the other discussions in a
voluminous literature on the subject see
\citet{Simpson49,Raup91,AdamiOfriaCollier00,Radick00}.

For the present purpose, we need to emphasize that, while the
issue of fundamental determinism or indeterminism is a
metaphysical one, in practice even perfectly deterministic
processes (like asteroidal motions or Buffon's matchsticks) are
often successfully modeled by stochastic methods. Even the fervent
supporters of convergence admit that there is much variation
between the actual realizations of the firmly fixed large-scale
trends, allowing a lot of margin for stochastic models. In the
context of researching SETI targets, for instance, the relative
difference in timescales of $10^6 - 10^7$ yrs makes quite
different accounts, although it could be argued that it is just a
small-scale perturbation or straying from the broadly set
evolutionary pathway. Recent studies, such as the one of
\citet{Borkowski09}, show that macroevolutionary trends
on Earth can be successfully described exactly within the
framework of the cellular automata models.

\subsection{Stepwise change in evolution}
Carter \citep{Carter83,Carter08}, \citet{Hanson98},
\citet{KnollBambach00} and other authors emphasize a number of
\textit{crucial steps\/} necessary for noogenesis. Some of the
examples include the appearance of the "Last Common Ancestor",
prokaryote diversification, multicellularity, up to and including
noogenesis. These crucial steps \citep["megatrajectories" in
terms of][]{KnollBambach00} might not be intrinsically
stochastic, but our present understanding of the conditions and
physico-chemical processes leading to their completion is so poor
that we might wish to start the large-scale modeling with only
broadly constrained Monte Carlo simulations. Subsequent
improvement in our knowledge will be easily accommodated in such
a framework (see also subsection \ref{frame} below). This applies
to any list of such steps (the problem, as Carter emphasized, is
that the number of really critical steps is quite controversial).
The work of \citet{PerezMercader02} shows how scaling laws can be
applied to the problem of the emergence of complexity in
astrobiology; in a sense, the present study is continuation and
extension of that work.

It is important to understand two different senses in which we
encounter stepwise changes in the astrobiological domain. In one
sense, we encounter models of punctuated equilibrium attempting
to explain the discrete changes in evolution, including possibly
catastrophic mass extinctions \citep[e.g.,][]{Bak97}. On the other
hand, megatrajectories can be generalized to cases in which we
are dealing with intentional actions, such as those which are
interesting from the point of view of SETI studies. Both of them
highlight the advantage of the discrete models like PCA over some
of the numerical work published in the literature, usually in the
context of SETI studies. For instance, \citet{Bjork07} calculates
the rate of colonization of planetary systems in the Galaxy under
relatively restricted conditions. This approach, pioneered by
\citet{Newman81} uses just a small part of the possible space of
states regarding capacities of advanced technological evolution.
If, as warned by the great historian of science, Steven J.~Dick,
postbiological evolution is the dominant general mode of
evolution in the last megatrajectory \citep{Dick03,Dick08}, many
of the concerns of SETI models based on continuous approximations
become obsolete \citep[see the criticism
in][]{CirkovicBradbury06}. On the other hand, stepwise changes
and phase transitions are generic features of a large class of
PCA \citep{PetersenAlstrom97}.

\subsection{Important global tendencies and redundant local
information} This methodological proviso -- historically the
all-important motivation behind von Neumann's introduction of
stochastic models in physics -- provides a rationale for similar
simulations in other fields of life sciences, in particular
ecology \citep{SoaresFilhoCerqueiraPennachin02} or epidemiology
\citep{HoyaWhite06}, or even urban traffic
\citep{BattyCouclelisEichen97}. Even more to the point of the
specifics of astrobiology, PCA have recently been successfully
used in the "Daisyworld" models \citep{WoodAcklandLenton06},
which share much of the complexity of the models of the GHZ
described below. We do not need to know specific details of
biogenesis, noogenesis and other processes on a particular planet
in GHZ in order to get a global picture of the GHZ evolution and
argue for or against particular research programs, for instance,
for or against a specific SETI targeting project.

This applies to temporal, as well as the spatial scales. Research
on both past and future of the universe (classical cosmology and
physical eschatology) demonstrates clearly defined timescales,
which could be treated as discrete units. Even lacking the
detailed information on the GHZ census at any particular epoch,
we might still wish to be able to say something on the overall
tendencies up to this day and into the foreseeable future. This
is analogous to the cases in which global tendencies of complex
systems are sought with evolutionary computation algorithms
\citep{deOliveira02}.

All this should be considered in light of the breakdown of the
long-held "closed-box view" of evolution of local biospheres of
habitable planets. In both astrobiology and the Earth sciences,
such a paradigm shift toward an interconnected, complex view of
our planet, has already been present for quite some time in both
empirical and theoretical work. In particular, possible
influences of Earth's cosmic environment on climate
\citep{CarslawHarrisonKirkby02,Pavlov05}, impact catastrophes
\citep{ClubeNapier90,Clube92,Asher94,Matese96,Matese98},
biogenesis \citep{Cockell00,Cockell03}, or even biotic transfer
\citep{Napier04,Napier07,WallisWickramasinghe04,www08} have become
legitimate and very active subjects of astrobiological research.
Thus, it is desirable to be able to consider them within an
integrative view, assigning them at least nominal quantitative
values, to be substituted by better supported data in the future.

\subsection{Framework adaptable to future observations and
results} \label{frame} PCA models in general rely on input matrix
of probabilities (of transitions between internal states). This
makes such models a very flexible tool, since such a matrix can be
fitted to any number of future observations, as well as conceptual
innovations and theoretical elaborations. In particular, it is to
be expected that on-going or near-future space-based missions,
like DARWIN \citep{Cockell09} or GAIA \citep{Perrymanetal01}, will
provide additional constrains on the input matrix of
probabilities. The same applies to future theoretical
breakthroughs, for instance the  detailed modeling of the
ecological impact of intermittent bursts of high-energy cosmic
rays or hard electromagnetic radiation. This will be accompanied
by "fine-graining" of the automaton states and of the network of
transitional probabilities.

\subsection{Historically used probabilistic arguments in SETI
debates} Many arguments used in SETI debates have been based on
probabilistic reasoning, the most important being the "anthropic"
argument of \citet{Carter83}. For elaborations on the same topic
see \citet{LineweaverDavis02,Davies03,CirkovicVukoticDragicevic09},
etc. Remaining in this same context offers clear advantages in
being able to account for various phenomena suggested as dominant
and get a historical perspective to this extremely rich
discussion. While this is a general argument for applying the
entire class of Monte Carlo simulations to astrobiological
problems, something could be said for the particular PCA
implementations of numerical models. The wealth of existing
knowledge on different PCA applications is immensely useful when
approaching a manifestly complex and multidisciplinary field such
as astrobiology and SETI studies.

Actually, it might be interesting (and historically sobering) to
notice that one of the fathers of evolutionary theory, Alfred
Russel Wallace, was a forerunner of astrobiology. He has actually
argued for the uniqueness of the Earth and humankind on the basis
of the cosmological model in which the Sun was located near the
center of the Milky Way similar to the long-defunct Kapteyn
universe \citep{Wallace03}. This remarkable, although incorrect,
argument demonstrates how important astrophysical understanding
has been since the very beginning of scientific debates on life
and intelligence elsewhere.

The SETI debate has, in the course of the last 4 decades, been
dominated by analysis of the Drake equation (Drake 1965), which
in itself is the simplest general probabilistic framework for
analysis of worthiness or else of SETI projects. Many criticisms
have been raised of the Drake equation
\citep[e.g.,][]{WaltersHooverkotra80,Wallenhorst81,Cirkovic04,Burchell06},
accompanied by suggestions of modification, but the key problem
remained: overall, the level of astrobiological numerical
modeling has remained non-existent to very low, so there has been
no viable alternative to the crudeness of the Drake equation. With
vastly widened spectrum of astrobiological research in the last
decade and a half, it seems appropriate to overcome this
deficiency and offer a new probabilistic framework in this area.
For a recent attempt along these lines see
\citep{Maccone10,Maccone12}. The PCA formalism we develop here has
the same essential form as the Drake equation: it uses a list of
input probabilities in order to generate a global conclusion
about the number and density of plausible observational SETI
targets. However, it adds much more information and can
incorporate many additional phenomena, like biotic feedbacks,
interstellar panspermia, etc.

\subsection{Practicality in parallelization}
Astrophysical numerical models are usually quite expensive in
terms of CPU time. Efficient parallelization is, therefore, not a
luxury but a necessity. CA models are particularly suitable for
this, since the non-parallelizable fraction (instructions dealing
with the state change of a single cell and its immediate
environment) is a very small part of the whole, and hence,
according to Amdahl's law and its modern multicore versions
\citep[e.g.,][]{Hill08}, the net speed gain is large.

\section{Probabilistic model of the GHZ}
The basic quantity, associated with the cell state in our PCA
models, is astrobiological complexity. It need not only describe
the complexity of life itself but additionally  the amount of
life-friendly departure from the simple high temperature/entropy
mixture of chemical elements commonly found in stars. The stars
can be considered as objects with zero astrobiological
complexity, while e.g., molecular clouds (hosts of complex
organic molecules) can be assigned an astrobiological complexity
slightly higher than zero. Further up this astrobiological
"entropy" scale, when it comes to planets, there is increasing
importance in considering the environment where the planet resides
and not just its intrinsic chemical composition. The sites with
the highest astrobiological complexity are life-bearing and
further quantified by the complexity of the life they host.

\begin{figure}
\centerline{\includegraphics[width=14cm]{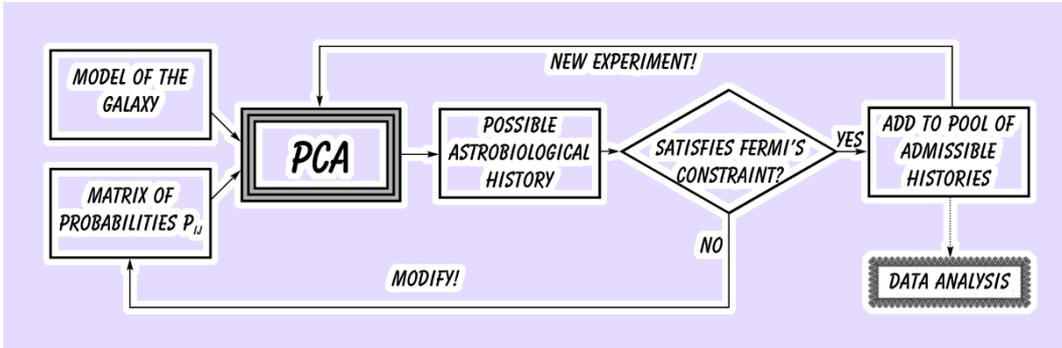}} \caption{A
scheme for modeling of the Galactic astrobiological evolution,
with a particular goal of resolving Fermi's Paradox in a specific
quantitative manner. Astrophysical model of the Milky Way, as
well as a particular choice of transition probabilities are input
data for the PCA kernel generating random possible
astrobiological histories, which can subsequently be tested for
any chosen boundary conditions, including those following from
Fermi's Paradox. Other specific versions of the same
philosophical approach are possible.} \label{pic1}
\end{figure}

We consider a probabilistic cellular automaton where cells of
four types occupy the sites of a regular square lattice of
dimensionality $D=2$ . Representing the Galaxy, especially the
astrobiologically interesting thin disk component, by a planar
system is physically justified \citep[see e.g.,][]{Binney87}, and
offers significant computational advantages over the realistic
$D=3$ equivalent models; we shall return to the relaxation of
this assumption in the final section. We model GHZ as the annular
ring between $R_{\rm inn} =6$ kpc and $R_{\rm out}=10$ kpc. We
use the absorbing boundary conditions at the boundaries, which is
rather obvious for modeling sites with simple lifeforms and
remains a good starting approximation for other cases. For the
purpose of simplification in our model development we have
adopted a discrete four state scale:
\begin{equation}
\sigma (i,j) = \left\{ \begin{array} {r@{\quad:\quad}l}
0 & \textrm{no life}\\
1 & \textrm{simple
life} \\
2 & \textrm{complex life} \\
3 & \textrm{technological civilization (TC)} \\
\end{array} \right. .
\end{equation}
The states and their considered transitions are shown
schematically in Figure 2. Of course, this is a very coarse
representation of the astrobiological complexity. The major reason
we believe it to be a good start for model-building is that these
scale points characterize the only observed evolution of life,
deduced from the terrestrial fossil record. The state $\sigma =1$
is, for instance, exemplified by terrestrial prokaryotes and
archea, while $\sigma =2$ is exemplified by complex metazoans.
The major model unknowns, i.e., the transition rules that model
the emergence and evolution of cells on the CA lattice are
implemented via probability matrix ($\hat{P}$). The relevant
biological timescales are directly modeled as the elements of
$\hat{P}$. With this approach various evolutionary scenarios can
be easily implemented by just changing the probability matrix
with no changes in model mechanics. We set the time step to be
equivalent to $10^6$ yrs (1 Myr).

Elements of:
\begin{equation}
\hat{P}^\mathrm{t}=\left( \begin{array}{ccccc} \vspace*{2mm}
P_{0,0}^\mathrm{t} &P_{0,1}^\mathrm{t}& \hdots &P_{0,m-1}^\mathrm{t}\\
\vspace*{2mm}
P_{1,0}^\mathrm{t} &P_{1,1}^\mathrm{t} &\hdots &P_{1,m-1}^\mathrm{t}\\
\vspace*{2mm}
\vdots\\
P_{m-1,0}^\mathrm{t} &P_{m-1,1}^\mathrm{t} &\hdots &P_{m-1,m-1}^\mathrm{t}\\
\vspace*{2mm}
 \end{array}
\right)
\end{equation}
are indicative of possible cell transitions for the m-state
(states range from 0 to m-1) PCA. In general, the state of the
cell in the next time step will result from the temporal
evolution of the cell itself (intrinsic evolution), influence of
the surrounding cells (local forced evolution) and Galactic
environment (externally forced evolution). For the above scaling
the full implementation of the transition probabilities can be
achieved with $4\times4\times6$ matrix\footnote{As for the
computationally more expensive models, where cell states take
continuous values from a predefined interval, the dimension of
probability matrix will influence the accuracy of interpolation,
when it comes to extraction of transition probabilities in such
models.} ($\hat{P}_\mathrm{ijk}$, where each k-part of
$\hat{P}_\mathrm{ijk}$ is represented by one matrix of
$\hat{P}^\mathrm{t}$ type). The $k=0$ part of $\hat{P}$ are $i
\rightarrow j$ transitions probabilities for intrinsic evolution
while the $k=5$ part are externally forced transition
probabilities. The rest of the matrix $\hat{P}$ describe the
forced evolutionary influence of the surrounding cell in state
$k-1$ on the $i \rightarrow j$ transition  for the cell in
question. Once developed PCA kernel is thus a highly adaptable
platform for modeling different hypothesis by just changing the
elements of $\hat{P}_\mathrm{ijk}$ as input parameters. There are
84 transition probabilities in total but most of them are likely
to be of no practical importance (e.g., colonization of a planet
by complex life -- state 2) or technically redundant (e.g., all
probabilities with $i=j$).

\begin{figure}
\centerline{\includegraphics[width=14cm]{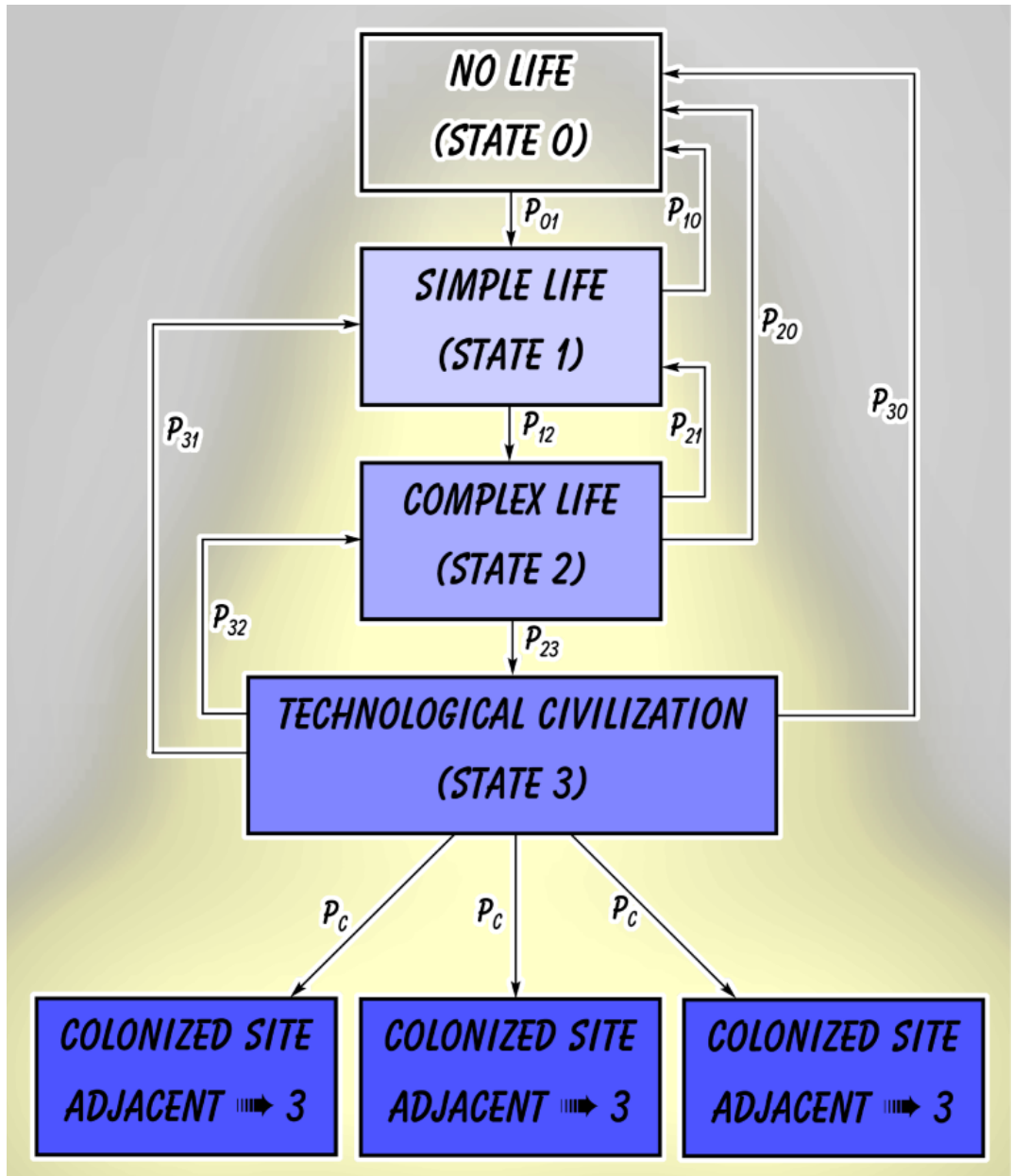}}
\caption{Steps in our PCA model reflecting major astrobiological
stages for evolution of each cell. In this scheme we have
neglected the possibility of interstellar panspermia, while the
possibility of panspermia within the same planetary system is
reflected in the increase in weight of each particular cell.}
\label{pic2}
\end{figure}

With the previously considered argument for discrete matter
distribution, it is likely that the majority of forced
probabilities can be disregarded. If we are to consider a cell on
the PCA lattice as a planetary system, truly significant effects
are likely to come from the $P_{034}$, $P_{134}$ and $P_{234}$
elements, that are indicatives of a TC colonizing the adjacent
sites. Also, the panspermia probabilities ($P_{012}$, $P_{013}$
and $P_{014}$), could be of some importance in denser parts of the
Galaxy, where the ratio of the average distance between the
adjacent planetary systems and an average planetary system size is
reduced. However, this is highly questionable since the more
populated parts of space experience the greater dynamical
instability which could seriously affect the habitability of the
comprised systems. The externally forced probabilities $P_{105}$,
$P_{205}$, $P_{305}$, $P_{215}$, $P_{315}$ and $P_{325}$ are
likely to be of greater importance. They are indicatives of the
global Galactic regulation mechanism (gamma-ray bursts,
supernovae, collisions, etc.) that are dependent on global
Galactic parameters (mainly star formation rate and matter
distribution) and can significantly alter the evolution of life.
However, these probabilities are still being strongly debated and
their influence on potential biospheres is somewhat
controversial. With the aforementioned probabilities being of
possible significance, the intrinsic probabilities are likely the
most important since they reflect the internal conditions in
planetary systems and on the planets themselves. Table
\ref{probabilities} lists the probabilities of possible
importance with a short description.

\begin{table}
\begin{center}
\caption{A list of significant probability matrix elements with a short description.}
\label{probabilities}
\small
 \begin{tabular}{p{0.1\textwidth}p{0.1\textwidth}p{0.7\textwidth}}
\hline
\hline
type$^a$&element&description\\
\hline
I&$P_{010}$&emergence of life in life friendly conditions\\
I&$P_{100}$&sterilization by increasing intrinsic life hostility (supervolcanism, biosphere collapse, parent star flares, asteroid impact, climate change, etc.)\\
I&$P_{120}$&evolution of simple to complex life under favorable conditions\\
I&$P_{200}$&similar as for $P_{100}$\\
I&$P_{210}$&extermination of complex life forms for similar reasons as in $P_{100}$\\
I&$P_{230}$&rise of the TC from complex life\\
I&$P_{300}$&sterilization caused by the reasons for $P_{100}$ plus TC induced reasons\\
I&$P_{310}$&extermination of complex life forms for reasons as in $P_{300}$\\
I&$P_{320}$&destruction of TC -- similar  as in $P_{310}$\\
F&$P_{034}$&TC expansion on adjacent systems -- colonization\\
F&$P_{134}$& the same as for $P_{034}$\\
F&$P_{234}$&the same as for $P_{034}$\\
F&$P_{012}$&panspermia\\
F&$P_{013}$&panspermia\\
F&$P_{014}$&panspermia\\
EF&$P_{105}$&sterilization by global regulation mechanisms (gamma-ray bursts, supernovae, etc.)\\
EF&$P_{205}$&the same as for $P_{105}$\\
EF&$P_{305}$&the same as for $P_{105}$\\
EF&$P_{215}$&extermination of complex life from the similar reasons as in $P_{105}$\\
EF&$P_{315}$&the same as for $P_{215}$\\
EF&$P_{325}$&destruction of a TC from the similar reasons as in $P_{105}$\\
\hline
\end{tabular}
$^a$ -- \footnotesize{I (intrinsic probability), F (forced probability), EF (externally forced probability).}
\end{center}
\end{table}

For some of the probabilities in the model, a simple
generalization of the known terrestrial conditions is possible. In
particular, this is the case with parameters $P_{010}$,
$P_{120}$, and $P_{230}$. The studies of the Earth's fossil record
have established the following timescales of ~1 Gyr, ~3 Gyr and
~600 Myr, respectively (actually these prototype values are
somewhat more conservative than those taken directly from the
terrestrial record, since both simple life and observers have
appeared more quickly on Earth). Despite numerous past debates,
there is still no consensus about the influence of extinction
events on the overall evolution of the terrestrial biosphere
[Gould's "third tier of evolution", \citet{Gould85}]. Even the
biggest known extinctions during the Phanerozoic eon did not
degrade the astrobiological complexity of the terrestrial
biosphere to the stage preceding the Cambrian explosion. In fact,
these events could might as well act as a "evolutionary pump"
because they have opened new ecological niches to a certain
species \citep{WardBrownlee00}; mammals experienced rapid advance
on account of the extinction of dinosaurs, after the K-T event. At
present, only vague estimates of the relevant
probabilities/timescales can be used.

\begin{table}
\caption{Fiducial values of transition timescales $\tau_{ijk}$
corresponding to input value transition probabilities $P_{ijk}$.
\label{nhco}}
\begin{center}
\begin{tabular}{|l|c|c|c|c|c|l|}
\hline No. & i & j & k & $\tau_{ijk}$ [yr] & $\delta \tau_{ijk}$
[yr] & Comment
\\ \hline \hline
1 & 0 & 1 & 0 & $1.0 \times 10^9 $ & $1.0 \times 10^9 $ &
"Copernican" hypothesis on biogenesis
\\
2 & 1 & 2 & 0 & $3.0 \times 10^9 $ & $1.0 \times 10^8 $ &
"Cambrian explosion" timescale
\\
3 & 2 & 3 & 0 & $6.0 \times 10^8$ & $1.0 \times 10^8$ &
Noogenesis timescale
\\
4 & 2 & 3 & 4 & $1.1 \times 10^8 $ & $1.0 \times 10^7 $ &
Expansion timescales
\\
5 & 1 & 3 & 4 & $1.1 \times 10^8 $ & $1.0 \times 10^7 $ &
\\
6 & 3 & 1 & 0 & $2.0 \times 10^7 $ & $1.0 \times 10^7 $ &
\\
\hline
\end{tabular}
\end{center}
\end{table}

The other group of input probabilities (comprised of remaining
$\hat{P}$ elements in Table \ref{probabilities}) is not known
empirically, even for the terrestrial case (one is tempted to
state: fortunately enough). In particular, we do not know the
probability of complex life on Earth going extinct in the next
Myr---although we are justifiably curious to get at least a vague
estimate of that particular parameter. Variation of this input
parameter makes for one of the most interesting applications of
the presented model in the future; for now, we have used fiducial
values inferred from the analyzes of \citep{Rees03} and
\citep{BostromCirkovic08}.

\begin{figure}
\centerline{\includegraphics[width=13cm]{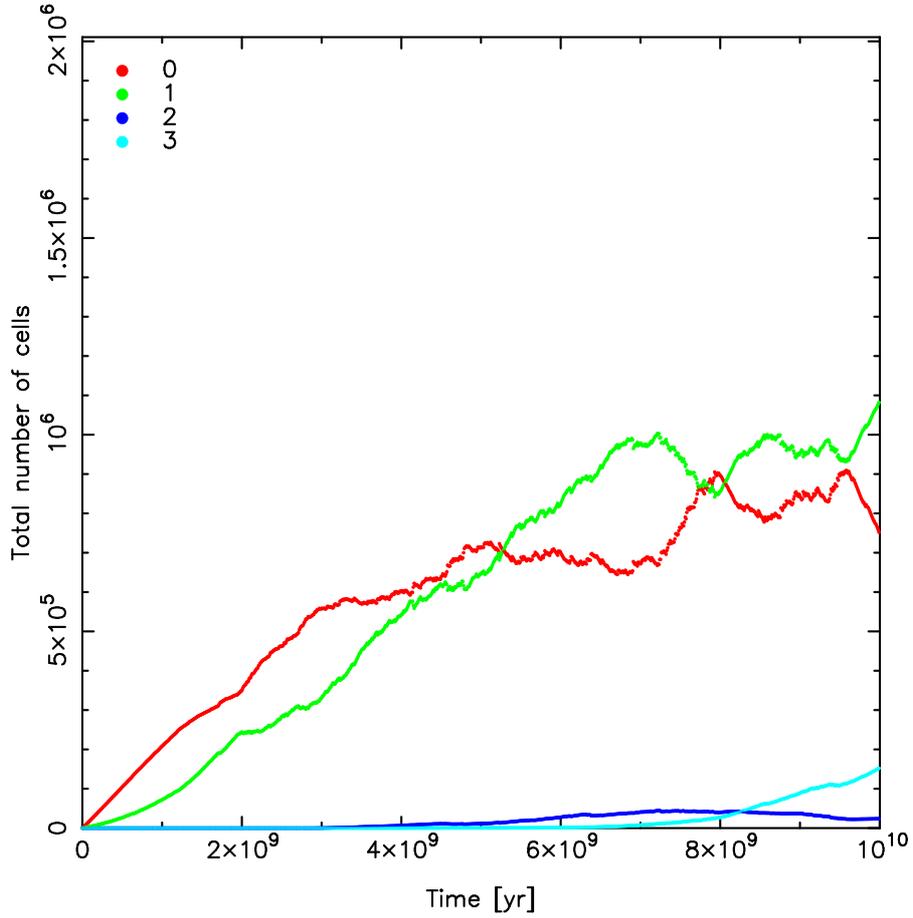}}
\caption{Evolution of populations of sites in various states
$i=0, 1, 2, 3$ (color-coded) in the PCA model of GHZ, averaged
over $N=10$ simulation runs. Timescale represents the age of the
thin disk of the Milky Way, corrected for the first 3.1 Gyr
lacking sufficient metallicity \citep{Lineweaver01}.} \label{pic25}
\end{figure}

Input distribution of Earthlike planet formation rate is given by
the seminal paper of \citep{Lineweaver01}. We use the
model of star-formation history of our Galaxy published by
Rocha-Pinto et al.\ \citep{Rocha-Pinto00a,Rocha-Pinto00b}. This is
more complex than the usually assumed quasi-exponential decay
form of star-formation density, but fits much better to
observational data on the age of populations, chemical evolution,
etc. We employ this form of star formation history as forcing the
evolution of Type II supernovae and gamma-ray bursts
\citep[astrobiological "reset" events; the choice of resets is
described in detail in][]{VukoticCirkovic08,Vukotic10}.

\begin{figure}
\centerline{\includegraphics[width=13cm]{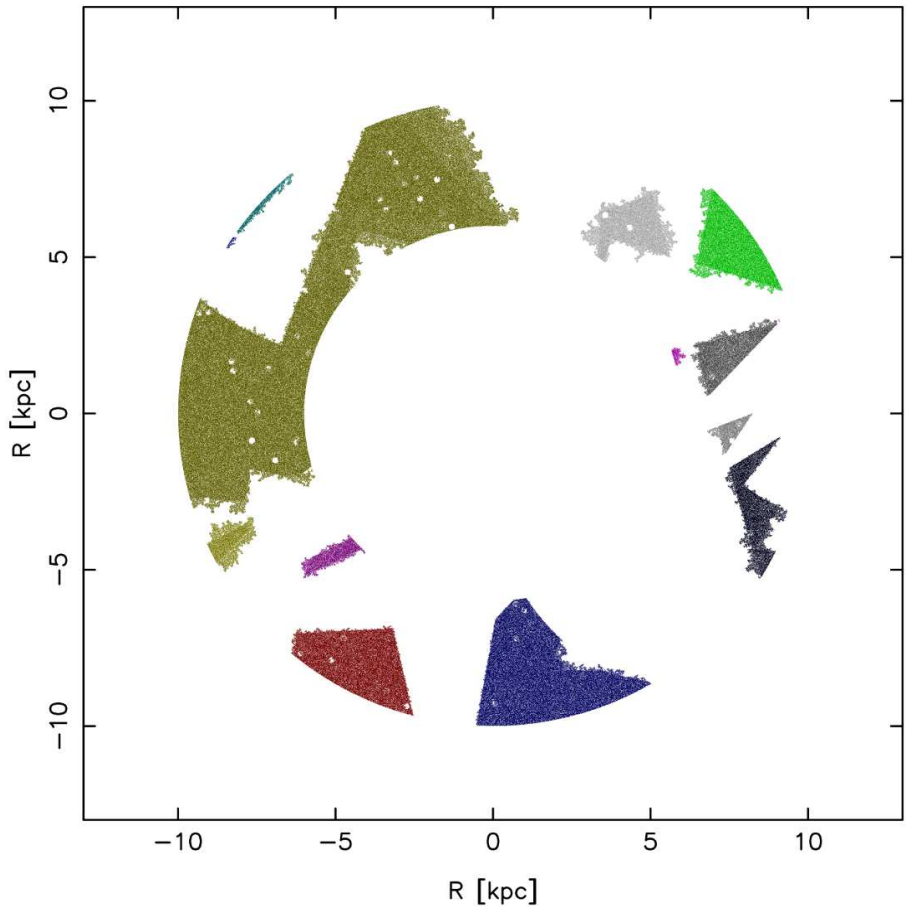}}
\caption{An example of clusters formed in the coarse-grained PCA
model of the Galactic Habitable Zone; scales are in kpc, and the
snapshot corresponds to "late" epoch.} \label{pic26a}
\end{figure}

After running $n= 10$ Monte Carlo simulations with synchronized
update at the spatial resolution of $R=100$ cells kpc$^{-1}$, we
analyze ensemble-averaged results. To get an overall picture of
the evolution of the system, we calculate the evolution of masses
in each state of our PCA:
\begin{equation}
\label{stanje1}
M_{\sigma} (t) = \langle \sum_{i,j} \delta_{c(i,j) \sigma} (t)\rangle,
\end{equation}
where
\begin{equation}
\label{stanje2}
\delta_{c(i,j) \sigma} \equiv \left\{ \begin{array} {r@{\quad:\quad}l}
0 & c(i,j) \neq \sigma \\
1 & c(i,j) = \sigma \\
\end{array} \right. ,
\end{equation}
is the Kronecker delta. This mass value counts different states at
each individual step $t$ and the average is taken over the number
of simulation runs. The results are shown in Fig.\ \ref{pic25}
averaged over $N=10$ runs, plotted against the age of Galactic
thin disk. We notice strongly nonlinear evolution, as well as the
numerical predominance of $\sigma =1$ cells at late times---which
can be construed as a support for the "rare Earth" hypothesis of
\citet{WardBrownlee00} \citep[see also][] {ForganRice10}. At $t
\sim 7000$ (corresponding to $\sim 3$ Gyr before the present)
first TCs appear in significant number, and the number of such
$\sigma=3$ sites increases subsequently \citep[non-monotonically,
though and not conforming to simple scaling relationship
occasionally suggested in the literature, e.g., in][]
{Fogg87,Bezsudnov10}. An example of the distribution of
$\sigma=3$ sites is shown in Fig.~\ref{pic26a}. (Lots of further
work needs to be done in order to highlight the sensitivity of
these results on individual input probabilities. A numerical error
at an early stage of the PCA kernel testing overestimated the
probability $P_{230}$ by about half of an order of magnitude in
comparison to the terrestrial value used here, accidentally
enabled us to test the sensitivity of the clustering analysis and
$V/V_0$, with encouraging results.)

\begin{figure}
\centerline{\includegraphics[width=13cm]{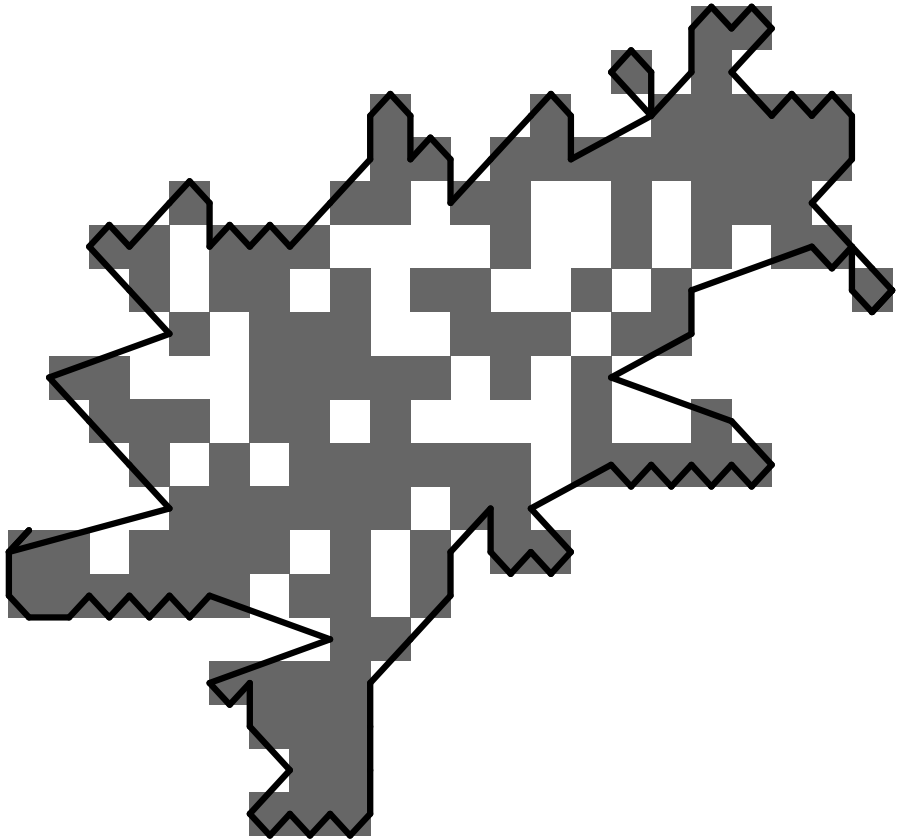}} \caption{An
example of the polygonal algorithm used for measuring the span of
clusters in the simulation.} \label{pic27}
\end{figure}

\begin{figure}
\centerline{\includegraphics[width=13cm]{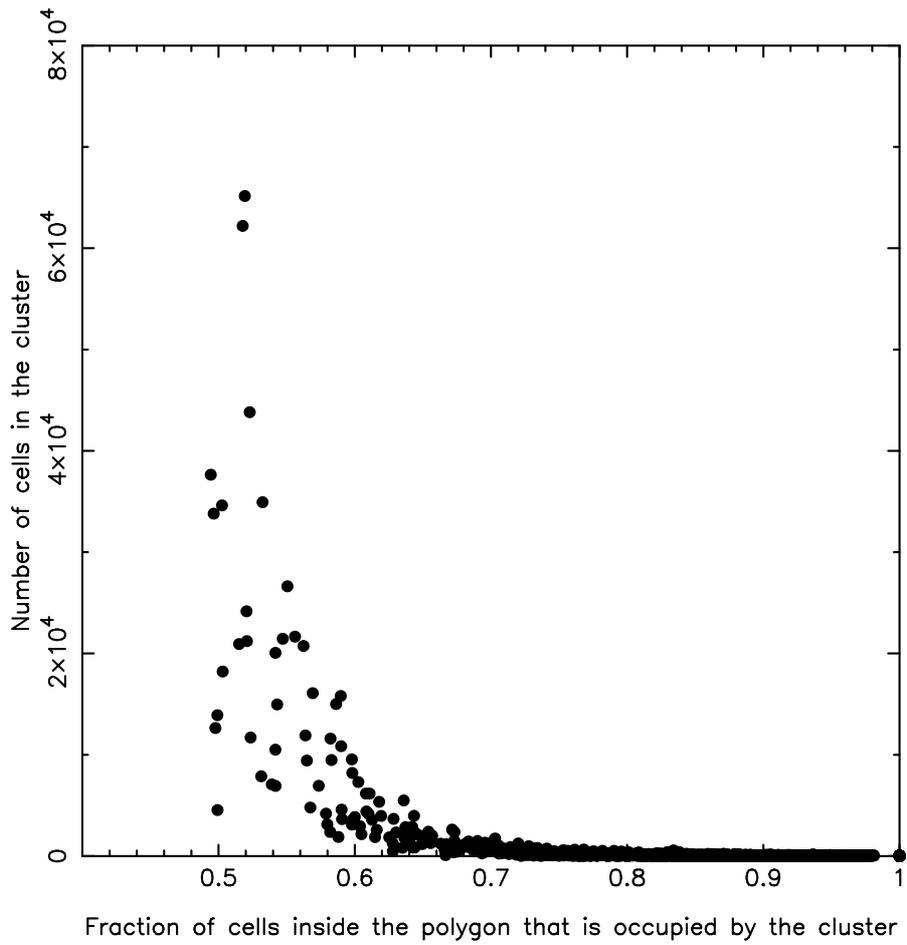}}
\caption{The distribution of mass filling factors of clusters of
state $\sigma =3$ ("technological civilizations") at epoch
$t=9500$, measured by the polygonal algorithm illustrated in
Fig.\ \ref{pic27}. Large clusters will tend to have filling
factors of $\simeq 50$\%, leaving many sites for continuation of
astrobiological evolution within their spans.} \label{pic30}
\end{figure}

In order to proceed with the analysis of clustering of such sites,
which is of obvious interest for practical SETI considerations,
we develop a polygonal representation of clusters, shown through
an example in Fig.~\ref{pic27}. Obviously (as in all forms of
percolation problem), clusters are porous structures, which may
contain many areas of persistence, suggested by \citet{Kinouchi01} as resolution of Fermi's Paradox. We test this
by investigating the fraction of cells inside the polygonal
representation that is occupied by the cluster at fiducial "late"
epoch of $t=9500$, by which we have at least one example of
spanning cluster in each simulation run, measured against the
total number of occupied cells within the cluster. (We use a
specific restricted sense of spanning cluster as the one which
spans the entire GHZ, that is has radial size of at least $R_{\rm
out} - R_{\rm inn}$, which seems appropriate for this particular
form of the percolation problem.) The results are shown in
Fig.~\ref{pic30} and are consistent with the distribution usually
obtained in clustering analyses of percolation in other contexts.
This serves as an auxiliary way of testing the proposed algorithm.

\begin{figure}
\centerline{\includegraphics[width=13cm]{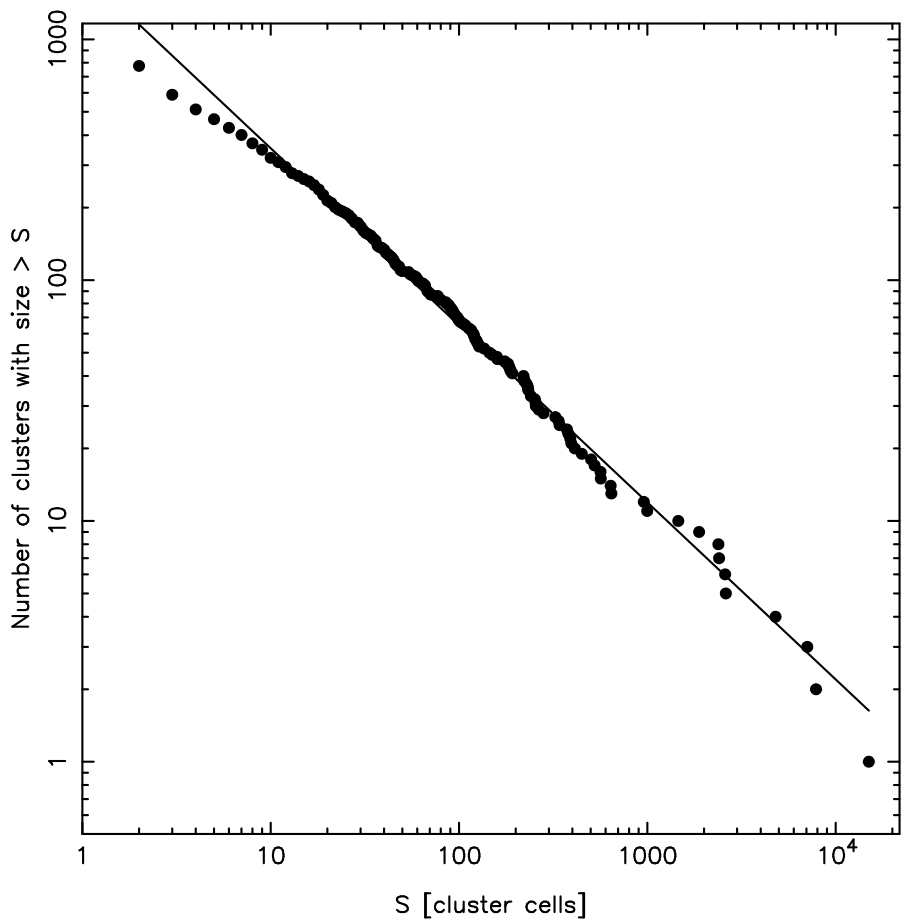}}
\caption{Mass index $\alpha= 1.72 \pm 0.01$ of the same set of
clusters as in Fig.\ \ref{pic30}.} \label{pic35}
\end{figure}

As shown in Fig.~\ref{pic35}, the set of clusters at the same
fiducial late epoch roughly obeys the scaling relation
\begin{equation}
N(>S) \propto S^{-\alpha},
\end{equation}
where $N(>S)$ is the number of clusters with more than $S$
cluster cells ("mass" of the cluster). The best-fit mass index is
given as $\alpha = 1.72 \pm 0.01$. Such behavior is remarkable in
view of the highly non-uniform underlying distribution of ages of
sites, represented by the planetary formation rate data and the
star-formation rate data influencing the distribution of the
reset events. The temporal dependence of this mass index in the
course of the Galactic history is shown in Fig.~\ref{pic54}, weak
increase in the last Gyr probably reflecting a sort of "natural
selection" favoring large clusters. While this might be an
important piece of information in debates surrounding, for
example, the famous Kardashev's classification of hypothetical
Galactic civilizations \citep{Kardashev64}, much further work is
required in order to better understand this behavior.

\begin{figure}
\centerline{\includegraphics[width=13cm]{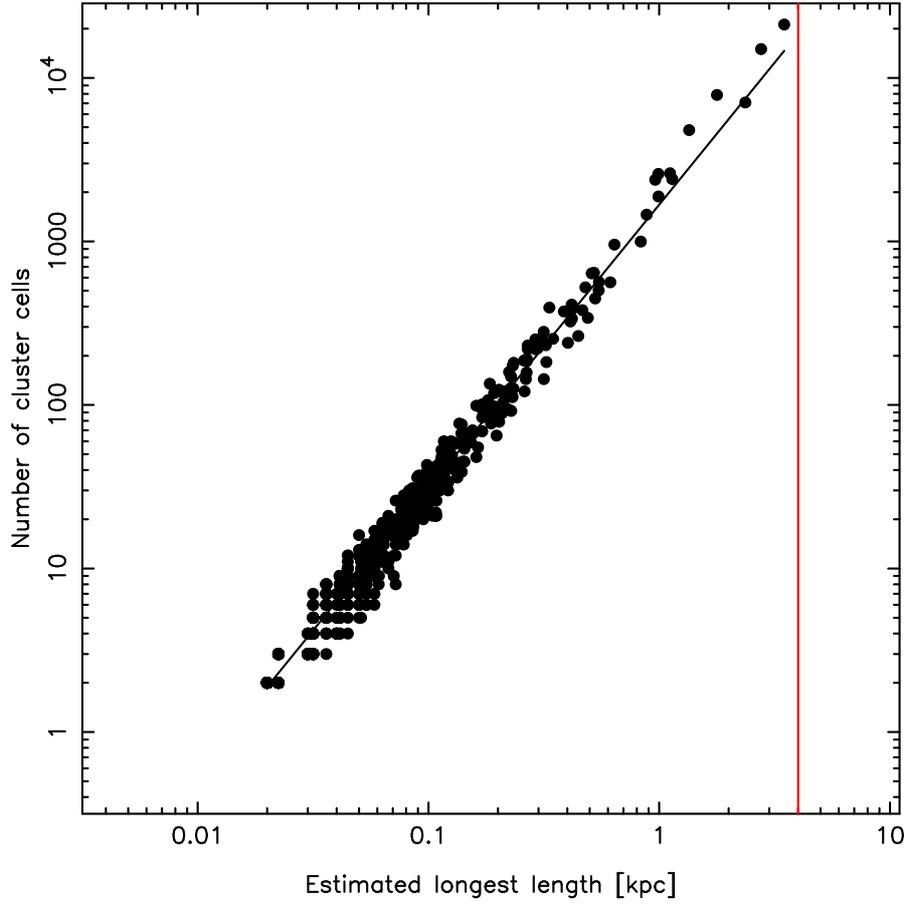}}
\caption{Lengthscale distribution of $\sigma =3$  clusters as
estimated by the polygonal method (see Fig.~\ref{pic27}).
Sparsely populated upper-right part of the diagram represents
what can be called "percolation" solution to Fermi's Paradox, as
suggested by Landis and Kinouchi (within the framework of our
neocatastrophic model, see text). Vertical line denotes the
radial size of GHZ, i.e., the quantity $R_{\rm out} - R_{\rm
inn}$.} \label{pic40}
\end{figure}

Finally, we need to consider the distribution of sizes of $\sigma
=3$ clusters, shown in Fig.~\ref{pic40}. Our results strongly
confirm the intuitive view that this distribution is strongly
time-dependent, on which most of the construals of Fermi's
Paradox are based. For the sample of chosen results---clusters at
$t=9500$---we notice that the highest concentration of clusters
is at $\simeq 0.1$ kpc, corresponding to small-to-medium sized
interstellar civilizations (for our, rather conservative, choice
of the colonization probabilities in the input probability
matrix), while the number of truly large clusters (equal or
larger to the size of GHZ itself) is marginal.

\begin{figure}
\centerline{\includegraphics[width=13cm]{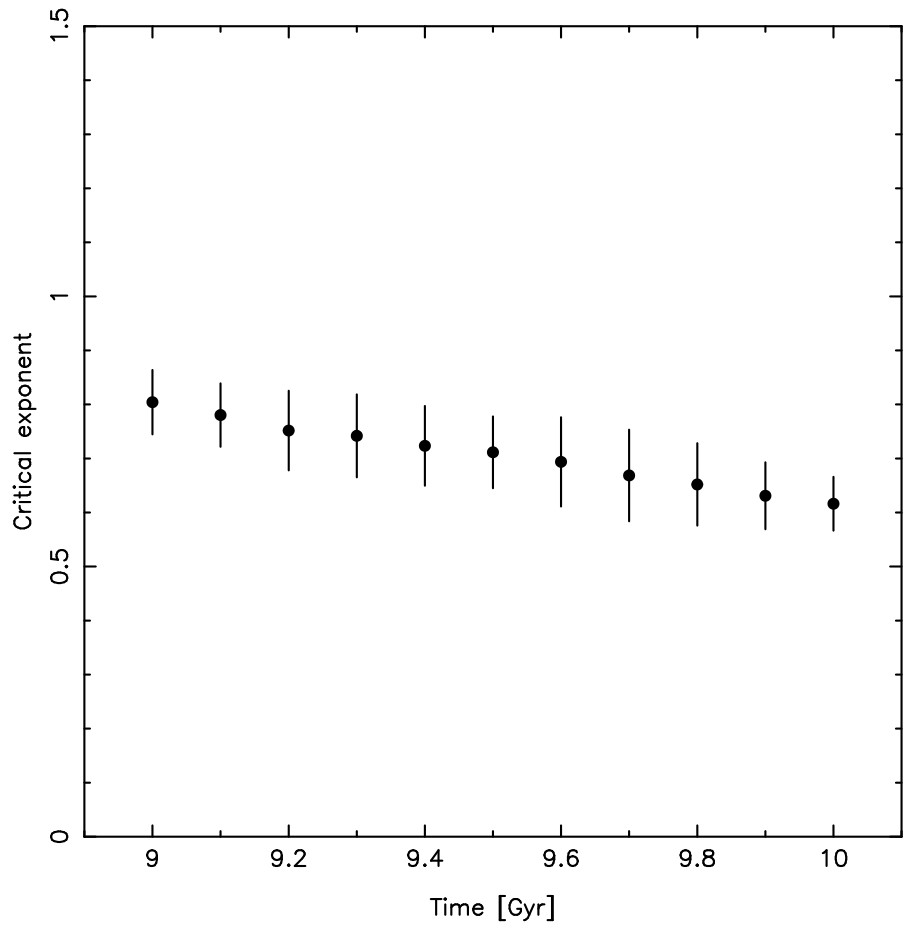}}
\caption{The evolution of critical exponent describing the set of
$\sigma =3$  clusters with time in the "late" epochs of the
history of astrobiological complexity of the Milky Way.}
\label{pic50}
\end{figure}

\begin{figure}
\centerline{\includegraphics[width=13cm]{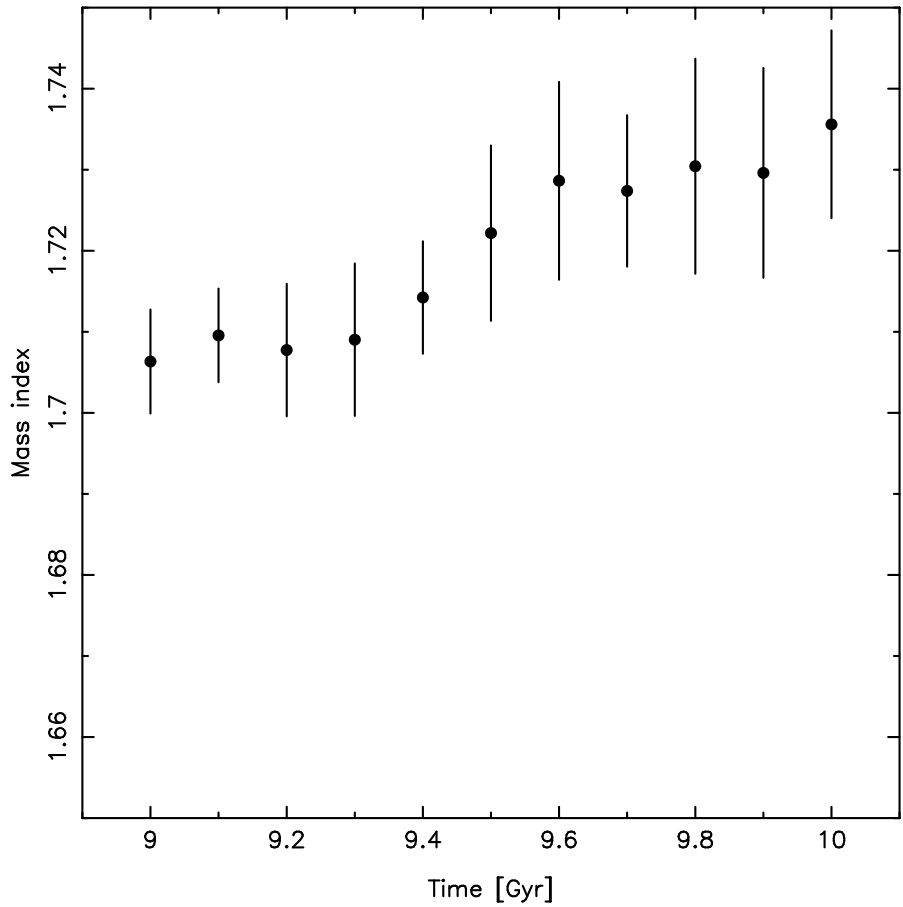}}
\caption{Behaviour of the mass exponent $\alpha$ for clusters of
state 3 ("advanced civilizations") shown during the last Gyr at
epochs separated by 100 Myr. Although within the probably
underestimated uncertainties, the rising trend is explicable as
those civilizations which survive tend to expand and add power to
the high-mass end part of the distribution.} \label{pic54}
\end{figure}

\begin{figure}
\centerline{\includegraphics[width=12cm]{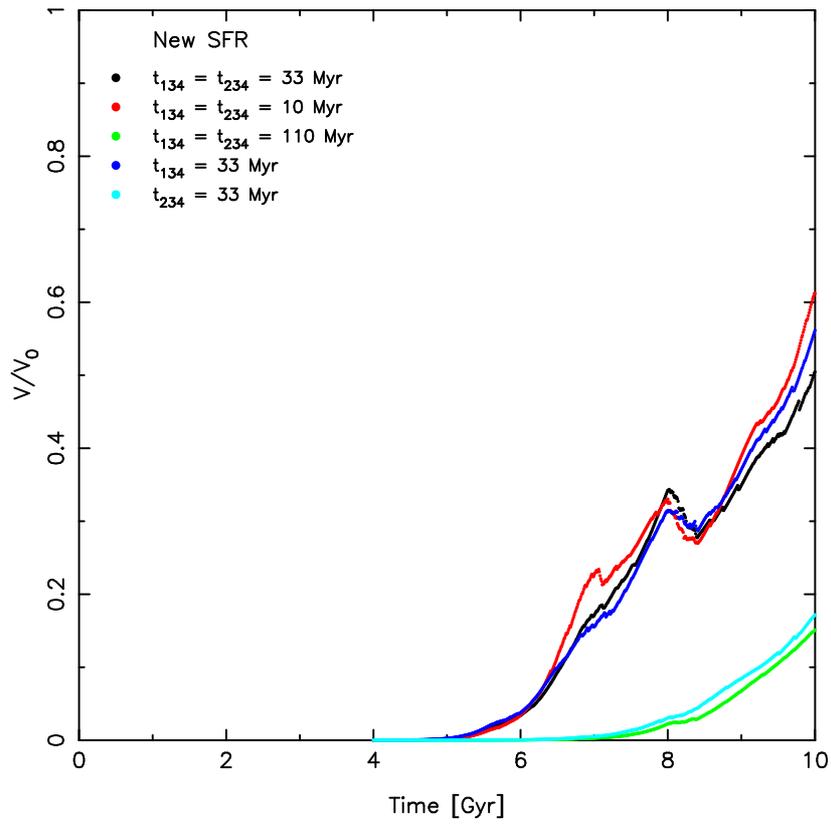}}
\caption{The average value of $V/V_0$ in the Milky Way during last
2 Gyr in our PCA model of GHZ. It is likely that we have
overestimated $V/V_0$ here and that this plot represents only the
lower limit, since the expanding wavefront of $\sigma =3$ sites
is likely to encompass sites outside of GHZ.} \label{pic60}
\end{figure}

In Fig.~\ref{pic60} we present the dependence of the relative
occupied volume in GHZ upon the time elapsed since the formation
of the Milky Way thin disk, averaged over 10 simulation runs for
various (color-coded) values of characteristic timescales. This
quantity, conventionally labeled $V/V_0$ (where $V$ is the
occupied volume, interpreted as the volume in which the presence
of a technological civilization is easily detectable) has
occasionally been used in SETI studies as a measure of ascent of
technological civilization on Kardashev's ladder. Here we have
used $V_0$ as the volume of GHZ in our $D=2$ model, suggesting
that we are in fact {\it overestimating\/} $V/V_0$, since it is
reasonable to assume that the expansion of technological
civilizations is not constrained in any significant manner by the
boundaries of GHZ. The important conclusion here is that although
we have started with "Copernican" input matrix of probabilities,
we still obtain $V/V_0 << 1$, which is in accordance with the
conventional interpretation of our observations related to
Fermi's Paradox. Thus, our PCA model seems to support the idea
that we can explain Fermi's Paradox in the framework of such
neocatastrophic discrete model. Although we have not run our
simulation far enough in the future (as defined by our chosen
timescale) to reach $V/V_0 \simeq 1$, it seems that the increase
is much shallower than in the models of \citet{Bezsudnov10}.

\section{Fermi's Paradox as a boundary condition}
The question of the astrobiological "landscape" of Galactic
evolution can be regarded as a particular instance of a (not
necessarily well-posed) boundary value problem. While we do not
understand the laws of local "astrobiological dynamics", we can
use boundary conditions, together with the assumption of the
local terrestrial example being randomly chosen from the
(unknown) distribution to constrain the space of possible
landscapes. Some of the boundary conditions are those we have
used in building of our PCA model: the age of the Galactic thin
disk, the boundaries of GHZ, the statistical distribution of
reset events. However, the most controversial one comes from
Fermi's Paradox \citep{Brin83,Webb02,Cirkovic09}.

In other words, can the famous lunch time question of Enrico
Fermi, "Where are they?", be helpful when it comes to answering
the question, "Where are we?" -- In what kind of neighborhood do
we exist? Depending on the aspirations of our possible "fellow
Galactizens" and the chances for their existence there are two
probable scenarios. For the purpose of this paper, we will
simply  called them soft and hard. The hard scenario puts more
weight on Fermi's paradox as a boundary condition, since it is
assumed that we have not yet observed an alien civilization
simply because there are no such civilizations capable \citep[or
willing, see][]{Cirkovic09} of interstellar travel and
communication. On the soft side, we can think of ourselves as
{\it being missed}, because we are residing in a passive pocket
of the Galaxy that is not near to any of the "highways" used by
other civilizations \citep{Kinouchi01}, or we are deprived of
contact of any kind \citep[the "Zoo hypothesis" of
Ball,][]{Ball73}. There are numerous assumptions that can be made
about the nature of Fermi's paradox---\citet{Smith09} concludes
that even after five decades there is still no way to find the
"right" values of the variables in the Drake equation, though it
is controversial for other reasons as well \citep[see
also][]{Cirkovic04}.

The hard version of the paradox will constrain the probability
matrix phase space that is indicative of sparse contact chances
throughout the Galactic history, while the soft version allows
for the phase space to be somewhat larger, meaning that there
were civilization contacts in the Galaxy but we just did not
experience them for various possible reasons. Obviously, the soft
version of the paradox acts as a more loose boundary condition
than the hard version. The porosity of large $\sigma=3$ clusters
in our simulations (Fig.\ \ref{pic30}), coupled with low $\langle
V/V_0 \rangle$ (Fig.\ \ref{pic60}), demonstrates how this still
seems acceptable within the "Copernican" framework, thus
essentially confirming the conclusions of \citet{Landis98} and
\citet{Kinouchi01}, but with addition of catastrophic reset
events. The downside of this is that one does not take into
account the fact that at least some of the manifestations of
advanced technological civilizations would be observable over
large interstellar distances \citep[e.g.,][]
{Freitas85,CirkovicBradbury06}. More research will be necessary
in order to quantify the conditions for such "Dysonian" approach
to SETI \citep{Dyson60,Sagan66,Carrigan09}.

Clearly, the issue will need to be settled by constructing a
whole series of models along the lines of our simple PCA, probing
large volumes of the input probability matrix space. Such
computationally more challenging programme will enable precise
determination of those chunks of parameter space consistent with
a particularly chosen form of Fermi's Paradox (for example, the
statement that there are no technological civilizations 1 Myr or
more older than us in the sphere of 100 pc radius around the Sun,
or a similar statement). This approach may be used as
complementary to the attempts to build a sounder theoretical
basis for SETI studies \citep{Maccone10}.

\section{Discussion and future plans}

We have analyzed a prototype 4-state astrobiological PCA whose
boundary conditions are derived from our understanding of
astrophysics and astrochemistry of the Milky Way, and whose
dynamical rules are inferred from our understanding of the
terrestrial biological evolution. It clearly belongs to Wolfram's
third class of cellular automata \citep{Wolfram83}, being able to
generate arbitrarily complex aperiodic states from a simple (in
our case even trivial) initial state. It is capable of generating
many possible astrobiological histories of our Galaxy, probing in
this way the huge parameter space involved. The main advantage of
the present approach is that the question "How many probable
solutions are there?" becomes for the first time numerically
tractable. By simply changing the values of $\hat{P}$ elements
over the part of the phase space of interest, we can model the
resulting astrobiological histories that have lead to the present
state. These parts of the phase space can be further interpreted
and connected with astrobiologically relevant processes and
events.

Even with the more restrictive hard version of Fermi's Paradox
there is still a great deal of $\hat{P}$ phase space to be
speculated about and included in the models. It would probably be
best to start with the smallest possible number of parameters.
The rest of the $\hat{P}$ elements can be considered in
subsequent phases of the iterative process in accordance with the
results of preceding simulations and their analysis. Instead of
implementing all elements listed in Table \ref{probabilities}, we
can restrict ourselves to implementing some of them or to subsume
a group of parameters into a single parameter.

A task of investigating the sensitivity on input parameters
remains; we can be the most comfortable with including the
elements related to at least an order of magnitude known
timescales ($P_{010}, P_{120}, P_{230}$). With the arguments of
the discrete matter distribution we can implement the forced
evolution with allowing only colonization by a neighboring cell
TC, such that $P_{034}=P_{134}=P_{234}$. Using the fact that, once
developed, complex life on Earth did not perish despite a few
major extinction events, it is probably justifiable to approximate
with  $P_{105}=P_{205}=P_{305}$ and with
$P_{100}=P_{200}=P_{300}$ (or perhaps separate $P_{300}$ from
some possible TC induced reasons, see \citet{BostromCirkovic08}.
Namely, once life reached the stage of advanced civilization it
is reasonable to assume that it cannot be easily degraded -- i.e.,
such a degradation could be possibly achieved with the sterilizing
disaster that will completely deprive the planet of living
organisms. With the exception of $P_{010}, P_{120}$ and $P_{230}$
there are three or four (with separate $P_{300}$) adjustable
parameters. By varying these parameters in our future
simulations, we can hopefully restrict their values to a narrower
range using Fermi's paradox as a boundary condition (cf. Duric
and Field 2003). Then a model could be further refined by
separating the equalities mentioned above.

Considering the vast uncertainties that characterize research of
this kind (some of them probably coming from the implicit
Copernican assumptions), despite the advantages of the approach
presented in this paper, we think that major improvements can be
made  with the incoming new data from future multidisciplinary
studies and space missions. In future work we are planning to
present and analyze the results of a similar PCA model with a more
detailed probability matrix, as well as higher spatial resolution
using massive parallel computing. Beside these, there are several
phenomenological improvements which seem to hold some prospects
for future work, deserving to be mentioned here.

Further improvement of boundary conditions can be implemented
with including colonization by TCs of sites beyond the boundaries
of GHZ \citep[in particular the large volume beyond $R_{\rm out}$
can be interesting for those advanced TCs motivated primarily by
optimization criteria,][]{CirkovicBradbury06}. An important
extension of the present model would be incorporation of
interstellar panspermia: the possibility of transfer of simple
lifeforms (commonly envisaged in form of extremophiles of {\it
Bacteria\/} or {\it Archaea\/} domains of life) from one
planetary system to another. Several viable theories have been
proposed recently
\citep{Napier04,Napier07,WallisWickramasinghe04}, whose common
property is that interstellar panspermia is very slow process.
Thus, we have not used it in obtaining the results presented
here, but the generalization is rather straightforward:
characteristic timescales are $\sim 10^9$ yrs for transfer
between neighbouring planetary systems at the Solar
galactocentric distance and correspondingly larger for more
distant systems, following roughly the random-walk arrival times.
This translates, in an ideal PCA with 1 pc-sized cells, into
interaction between neighbours, weighted by the mean stellar
density in our Milky Way model in the same way as the density of
planetary systems (as well as the density of supernovae/gamma-ray
bursts) is weighted. In more coarse-grained simulations,
panspermia would increase the biogenesis potential of a single
cell and possibly act to reduce timescale for transition to
complex life in it\footnote{This might be the case with the Solar
System as well---since there are several hypotheses of
biotic/prebiotic exchange between two or more Solar System
bodies, mainly contenders being early Mars and Earth \citep[cf.][]
{Burchell04,Davies03,Levin07}.} or to increase the sterilization
timescale (making simple life more persistent in a single cell of
the automaton). While it is hard to gauge the overall impact of
panspermia on the Milky Way astrobiological landscape, it is
conceivable that locally---including the neighbourhood of the
Solar System---and, in the long run, it may make a difference;
only detailed future work can resolve the issue. Beside
increasing spatial, one may strive to increase temporal
resolution as well, in particular when it comes to modeling of TC
clustering.

An additional refinement left for future models is taking into
account the differential rotation of the Galaxy. Such rotation
will cause continuous deformation of clusters on timescales $\sim
10^8$ years and larger. Since the kinematics of the Milky Way is
rather well-understood, it is conceptually straightforward to
apply this to our GHZ model, although the computational
implementation is, according to preliminary considerations,
rather expensive. It seems that this effect might be of some
importance in the late epochs for TC clusters of large span. It
has been intuitively suggested as a difficulty for those answers
to Fermi's paradox like Fogg's "Interdict Hypothesis"
 relying on large- scale uniformity of behavior of
TCs \citep{Fogg87}; further development of quantitative GHZ models
will present an opportunity to check this intuition numerically.

Finally, an obvious further step is building $D=3$ PCA models,
reflecting the vertical stratification of Galactic matter, as
well as some possible additional effects on local biospheres,
e.g., Galactic plane/spiral arms' crossings and their ecological
consequences \citep{Leitch98,Gies05}. This will add a new layer of
complexity and re-emphasize the degree in which local biological
conditions are embedded in wider and richer astrophysical
surroundings.

In conclusion, PCA seems to be a fruitful approach for the
quantitative approach to astrobiology in the Milky Way context.
Although still plagued by many uncertainties, quantitative
astrobiology has wide perspectives, utilizing the best of
computational physics of today, together with continuously
updated observational data from the new generation of
astronomical instruments. In particular, it carries the prospect
of at least better framing---if not answering---perhaps the most
intriguing question in all science, "Are we alone?" \vspace*{1cm}

\noindent{\bf Acknowledgements.} The authors use this opportunity
to thank Carlos Cotta, Anders Sandberg, Ivan Almar, Geoffrey
Landis, Branislav Nikoli\'c, Aleksandar Obradovi\'c, Jelena
Andreji\'c, Claudio Maccone, Zona Kosti\'c, Steven J. Dick, and
the late Robert Bradbury for their valuable input, kind support,
and technical help. The Editor, Alan W. Schwartz, has been
instrumental in vastly improving a previous version of the
manuscript. This research has been supported by the Ministry of
Education and Science of the Republic of Serbia through the
Project \#176021 "Visible and Invisible Matter in Nearby
Galaxies: Theory and Observations".

\vspace{1cm}
\fontdimen2\font=0.4em
\fontdimen3\font=0.16em
\fontdimen4\font=0.16em
\fontdimen7\font=0.2em
\hyphenpenalty=0
\bibliography{mybib}
\end{document}